\documentclass{article}

\usepackage{arxiv}

\usepackage[utf8]{inputenc} 
\usepackage[T1]{fontenc}    
\usepackage{hyperref}       
\usepackage{url}            
\usepackage{booktabs}       
\usepackage{amsfonts}       
\usepackage{nicefrac}       
\usepackage{microtype}      
\usepackage{lipsum}
\usepackage{graphicx}
\usepackage{multirow}
\usepackage{caption}
\usepackage{subcaption}

\usepackage{array} 

\graphicspath{ {./figures/} }

\title{CATBERT: Context-Aware Tiny BERT for Detecting Social Engineering Emails}

\author{
  Younghoo Lee \\
  Sophos AI\\
  \texttt{younghoo.lee@sophos.com} \\
   \And
  Joshua Saxe \\
  Sophos AI\\
  \texttt{joshua.saxe@sophos.com} \\
   \And
  Richard Harang \\
  Sophos AI\\
  \texttt{richard.harang@sophos.com} \\
}

\begin{document}

\maketitle

\begin{abstract}
Targeted phishing emails are on the rise and facilitate the theft of billions of dollars from organizations a year. While malicious signals from attached files or malicious URLs in emails can be detected by conventional malware signatures or machine learning technologies, it is challenging to identify hand-crafted social engineering emails which don’t contain any malicious code and don’t share word choices with known attacks. To tackle this problem, we fine-tune a pre-trained BERT model by replacing the half of Transformer blocks with simple adapters to efficiently learn sophisticated representations of the syntax and semantics of the natural language. Our Context-Aware network also learns the context representations between email’s content and context features from email headers. Our CatBERT(Context-Aware Tiny Bert) achieves a 87\% detection rate as compared to DistilBERT, LSTM, and logistic regression baselines which achieve 83\%, 79\%, and 54\% detection rates at false positive rates of 1\%, respectively. Our model is also faster than competing transformer approaches and is resilient to adversarial attacks which deliberately replace keywords with typos or synonyms. 
\end{abstract}

\keywords{Social Engineering Emails \and Phishing \and Machine Learning \and Natural Language Processing \and BERT}


\section{Introduction}
\begin{figure}
  \centering
  \includegraphics[scale=0.5]{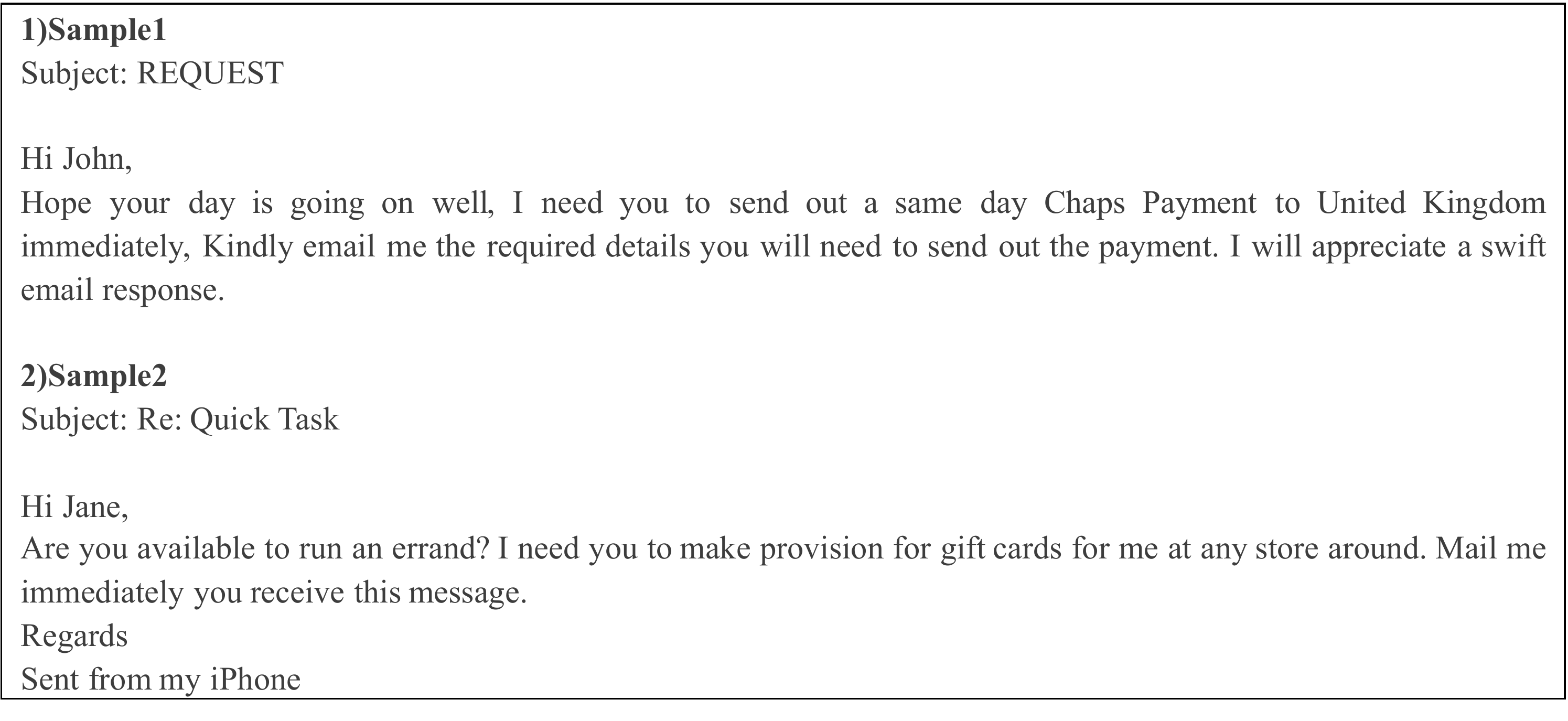}
  \caption{Examples of social engineering attacks.}
  \label{fig:bec_samples}
\end{figure}

Social engineering attacks leveraging hand-crafted emails are a major category of cyber crime. Because these emails are often hand-written, individually targeted, and frequently incorporate background research on their targets \cite{ho2017detecting}, they pose a significant challenge for conventional detection systems which rely on spam-like duplication between previously seen and new malicious emails for the identification of these emails. Indeed, targeted phishing emails may not share any word sequences or word choices with previously seen attacks and may appear different in only subtle ways from benign messages. Figure \ref{fig:bec_samples} lists two targeted phishing emails to illustrate the look and feel of hand-crafted phishing emails. While these samples do not contain any malicious attachments or phishing URLs, they compel victims to comply with requested behavior as part of attack playbooks. 

To detect targeted phishing emails, we fine-tuned a customized BERT-based model that has been pre-trained on billions of words of benign text to learn a sophisticated representation of the syntax and semantics of natural language \cite{devlin2018bert}. This allows the network to pick up on the subtleties of email topic, tone, and style.  Our fine-tuning approach then significantly reduces the complexity of the network and optimizes the network to detect phishing attacks. 
BERT is a Transformer-based natural language processing (NLP) model that has achieved great success in NLP tasks including sentiment classification, machine reading comprehension, and natural language inference. However, deploying full-scale BERT models, with up to 340 million parameters, for real-time security applications remains challenging. Our fine-tuning approach obtains a smaller and faster model with substantial performance improvement. In addition to the text data in emails, header fields provide additional contextual information about the relationship between senders and recipients. Our network architecture combines email’s content data with this additional context and improves the classification performance over vanilla BERT. 

The contributions of our work are listed below. 
\begin{itemize}
\item We propose a BERT-based phishing email detection model which learns sophisticated representations from the content data in email’s body and context data in email’s header. The proposed model outperforms a vanilla BERT model fine-tuned on the same training data. 
\item Full-sized BERT models are quite heavy for real-time security applications. Our fine-tuning approach substantially reduces the model’s complexity by replacing one out of two Transformer blocks with simple adapters and achieves a smaller and faster model. 
\item Our model is robust to adversarial attacks which deliberately replace words with typos or synonyms.    
\end{itemize}

The rest of the paper is organized as follows. Section \ref{sec:relatedwork} describes the related work. In section \ref{sec:proposedmethod}, we elaborate our proposed method and then discuss experiments in section \ref{sec:experiments}. The work is concluded in section \ref{sec:conclusion}.

\section{Related work}
\label{sec:relatedwork}

Our work relates to research in machine learning models for text classification and model compression methods for BERT models.

\subsection{Machine learning approaches}
Attackers have been using a variety of techniques to evade existing detection methods, which makes signature-based approaches less effective to detect ever-evolving attacks. Researchers proposed various machine learning models to solve the problem. Traditional NLP approaches, still widely used in practice, extract TF-IDF (Term Frequency - Inverse Document Frequency) features from word tokens and train logistic regression, tree-based and support vector machine (SVM) models \cite{ho2019detecting, bhowmick2016machine, cidon2019high}. More recent work has employed RNN (Recurrent Neural Network) or CNN (Convolutional Neural Network) models on sequential language data \cite{lukas2020catching, nikita2020deepquarantine}. 

The authors of \cite{ho2019detecting} trained a Random Forest classifier with features from ``phishy'' keywords and URL reputation information and discovered unseen phishing attacks from a large-scale dataset. While their method relies on historical statistics from a large dataset to collect reputation scores for URLs and calculate the similarity scores of recipients, the historical communication data is not always available. The work of \cite{lukas2020catching} proposed a LSTM(Long Short Tern Memory)-based model to detect phishing attacks using word sequences using a word vocabulary of 4995. Our experimental models also include a LSTM-based model with about 120,000 tokens from a Multilingual BERT tokenizer, which outperformed a non-sequence-based Logistic Regression baseline model. However, our experiments show that the LSTM model was outperformed by Transformer based models. In \cite{nikita2020deepquarantine}, a CNN-based model was introduced with features only from email headers to detect potential spam messages. Our experiments demonstrate that combined features both from email’s header and body improves the detection performance against various phishing attacks.

\subsection{Transformer models}
The model we propose here is based on a relatively new neural network construct called the transformer block.  Transformer blocks, attentional mechanisms that output context vectors for each word that depend on the context for that word, are the core component in self-attention language models such as BERT \cite{devlin2018bert} and GPT \cite{radford2019language}, and became well known when researchers demonstrated state-of-the-art results in 11 NLP tasks using BERT in 2018. 

Our model, which is derived from BERT, has a training procedure consisting of two stages. In the first pre-training stage, a (full-size) BERT model is trained to predict masked words in a sentence with large-scale datasets. The model also learns to correctly predict next sentence prediction. In the second stage, the pre-trained model is fine-tuned with a labelled target dataset for a specific classification task. For the fine-tuning step, a classification header is added and the whole network is jointly optimized with training labels. Since the model's parameters are pre-trained in the first stage with large-scale data, fine-tuning only requires training with a small labelled dataset for two or three epochs to yield state-of-the-art performance. 

OpenAI’s GPT, a differing transformer block-based approach, also achieved state-of-the-art results in numerous NLP tasks in 2018. The model uses a standard language modelling objective to maximize the likelihood of a next token for a given sentence. The model is comprised of multiple Transformer blocks which apply a multi-headed self-attention operation over the input text. After training the model by predicting the next work in a sentence, the model is further trained by both optimizing a supervised objective and a language modelling objective. While GPT can be fine-tuned for classification tasks, the best application for the autoregressive language model is text generation. In our preliminary experiments, both GPT and BERT models obtained comparable performance when they contain a same number of Transformer blocks.

\subsection{Model compression}
While complex BERT models obtain impressive results, the full-sized models are computationally expensive and memory intensive. Model compression methods have been introduced to solve the issues with large models by reducing the parameters of models without significantly decreasing the model performance. Parameter pruning \cite{han2015deep} and knowledge distillation \cite{hinton2015distilling} are efficient approaches for model compression. 

Pruning methods remove less important neurons or collections by measuring the importance of neurons. While these methods result in a smaller sparse network, the speedup of inference time is not guaranteed as many deep learning frameworks do not fully support sparse operations. 

Knowledge distillation methods compress deep networks into shallower ones where a compressed model, a student network, mimics the function learned by a complex model, a teacher network. One of the advantages of knowledge distillation is that any student architecture can be trained with a complex teacher network. The method trains a student model with a standard classification loss and an additional distillation loss. The distillation loss indicates the output differences between the two models and allows the student to learn rich representations from the large teacher. 

Several compressed BERT models, such as DistilBERT \cite{sanh2019distilbert}, and ALBERT \cite{lan2019albert} have been proposed. DistilBERT reduces the number of Transformer blocks and ALBERT reduces the model size by sharing a Transformer block in cross layers. Our approach is to reduce the number of Transformer blocks and replace missing ones with simple adapters which leads to more stable and improved results.

\section{Proposed method}
\label{sec:proposedmethod}
State-of-the-art NLP models such as BERT and GPT have achieved great success in Natural Language Processing (NLP) tasks including sentiment classification, machine reading comprehension and natural language inference. Their success stems from semi-supervised learning which makes use of both unsupervised and supervised approaches and efficient self-attention architecture. The state-of-the-are NLP models have self-attention layers which consist of Transformer blocks, and they outperform existing models including Multi-Layer Perceptron (MLP) and Recurrent Neural Network (RNN) models. Semi-supervised learning is a two-stage procedure including pre-training with a large unlabeled dataset and fine-tuning with a small labelled dataset. The learning procedure allows us to fine-tune a pre-trained BERT model for our email classification problem. In this section, we present our network architecture and fine-tuning method that improves the performance of runtime and classification.

\subsection{Network Architecture}
We fine-tune a pre-trained DistilBERT model which has been compressed from a standard BERT model. A student network (DistilBERT) was trained with a large teacher network (BERT) in pre-training and the thinner student model obtained comparable performance with its teacher network \cite{sanh2019distilbert}. Since the Transformer blocks in DistilBERT were pre-trained with large datasets, our approach is to further compress DistilBERT by reducing the number of Transformer layers. We replace the removed Transformer blocks with simple adapters. The adapter blocks are inspired by the idea from parameter-efficient transfer learning \cite{houlsby2019parameter} which adds tiny adapters insider of each Transformer . Their main motivation of inserting adapters was to reduce fine-tuning costs for many downstream tasks. Their fine-tuning method allows downstream models to reuse a large but fixed Transformer parameters with fine-tuned task specific adapters. However, their method increases the inference cost at downstream models as the adapters add additional computation cost for prediction at production systems. Our approach replaces some of Transform blocks with simple adapters which improves inference speed. We also fix the lower two Transformer blocks but fine-tune remaining upper Transformer blocks. In our partial fine-tuning experiments, we froze two lower Transformer blocks and jointly fine-tuned all adapters and one upper Transformer block with a downstream task.   

\begin{figure}
  \centering
  \includegraphics[scale=0.6]{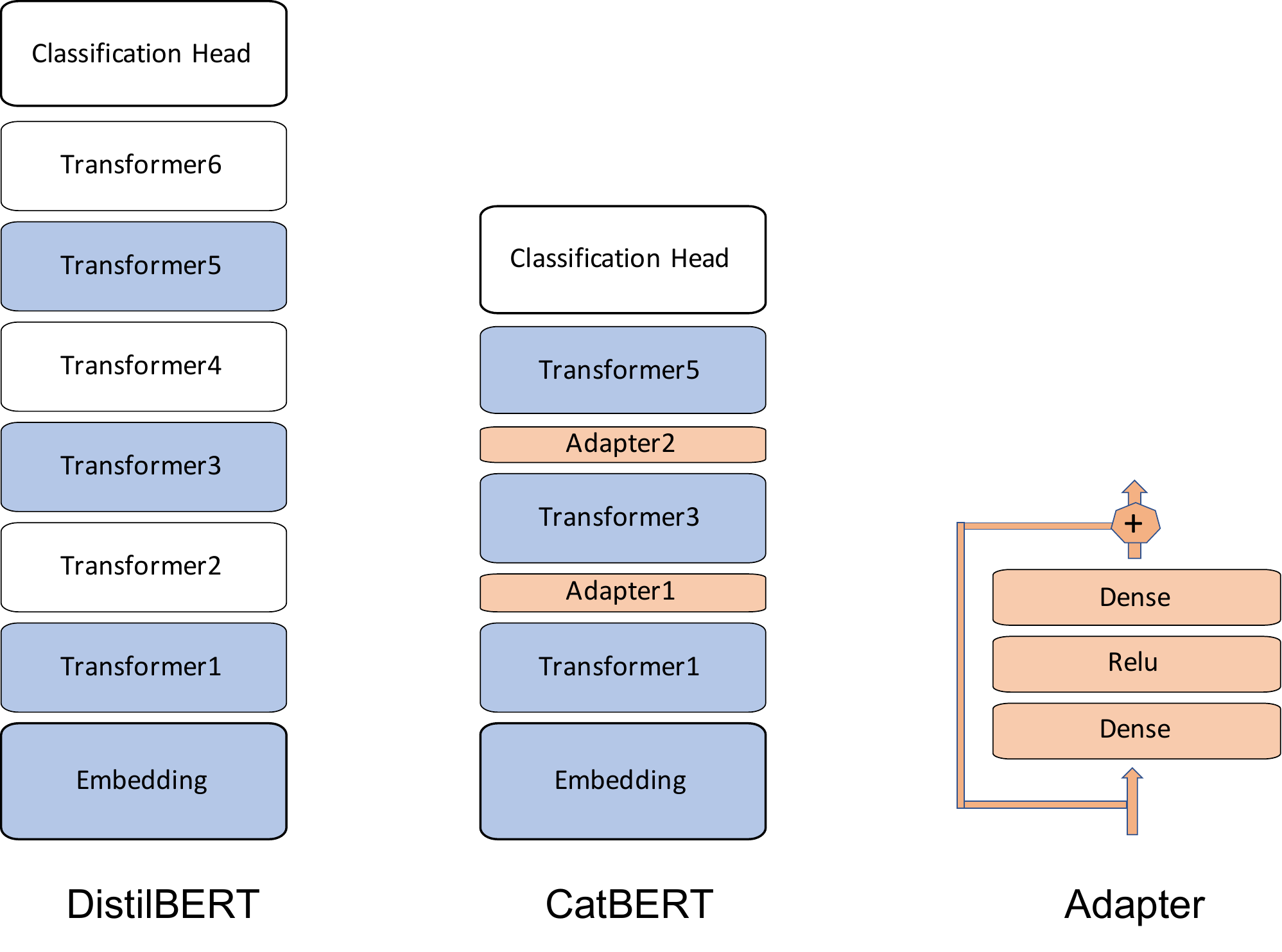}
  \caption{CatBERT is compressed from DistilBERT by taking odd-numbered Transformers and replacing missing Transformers with simple Adapters. The blocks in blue indicate that their parameters are initialized from DistilBERT. The Adapter layer consists of two Dense and a Relu activation units with a residual connection.}
  \label{fig:network_architecture}
\end{figure}

While DistilBERT reuses the same core network architecture used in BERT, the number of Transformer layers is reduced from 12 to 6 by taking one layer out of two. Our email model, CatBERT further compresses DistilBERT by a factor of 2. As DistilBERT was pre-trained with a large teacher using large-scale datasets, we reused the optimized parameters by initializing our slim BERT model from the DistilBERT, then removing half of the Transformer blocks. Our fine-tuning method directly optimizes the resulting network’s parameters without a teacher model because our preliminary experiment showed that Knowledge distillation on CatBERT did not improve detection performance, but did significantly increase the cost of training. Figure \ref{fig:network_architecture} depicts how we reduced the number of Transformer blocks. Simply removing every even-numbered or odd-numbered Transformers allows us to reduce the size and complexity of Transformer-based models by a factor of two. The first, third and fifth Transformers and the bottom embedding layer were initialized from the parameters from DistilBERT. However, the top classifier head was randomly initialized. The missing Transformer blocks are replaced with simple randomly initialized, trainable Adapter blocks. The Adapter block consists of a fully connected dense, a Relu activation and a second Dense units with a residual connection and the Dense units have the same dimensionality with Transformers.  Surprisingly, simply removing half of the transformer blocks and replacing them with trainable Adapter blocks was sufficient to surpass the high performance of the full model, despite the presumed interdependence between successive blocks in the original DistilBERT model. We do not explore this further but note it as an area for future study. 

We fit the network using a binary cross entropy loss function. The loss L is defined by given the output of our model $ f(x;\theta) $ for input $x$ and label $ y \in {0,1} $ and model parameter $\theta$.

\[ L(x,y;\theta) = -y \log(f(x;\theta)) + (1-y) \log(1-f(x;\theta))  \]

We solve for $\hat{\theta}$ the optimal set of parameters that minimize the loss over the dataset: 
\[ \hat{\theta} = \arg \min \limits_{\theta} \sum_{i=1}^{n} L(x_i,y_i;\theta) \]

Where N is the number of samples in our dataset, and $x_i$ and $y_i$ are the feature vector of the $i^{th}$ training sample and the label respectively.

\subsection{Content features}
The state-of-the-art NLP models preprocess raw text and convert the text to a sequence of word tokens. Traditional NLP models use a pre-defined vocabulary for tokenization, but recent models such as BERT and GPT employ sub-word tokenizers. We use BERT tokenizer which divides a complex word into simple sub-words and maintains a small vocabulary of 30,522 tokens for English models and 119,547 tokens for Multilingual models \cite{devlin2018bert}. The sub-word tokenizer can solve out-of-vocabulary problems, as unknown words are tokenized into sub-words. Token inputs are represented as token embeddings and the position information of tokens is represented as positional embeddings. There are two special tokens, CLS is the first token and SEP is the last token for every text input. BERT-based models can handle up to a maximum sequence length of 512. 

We truncate text from the beginning of tokens when the length of tokens is larger than the limit. The hidden state for CLS token from the last Transformer block is fed into the final classification layer. For example, when subject and body text are given, concatenated BERT tokens are provided as the input to token embeddings block and the position information of tokens is additionally fed into the position embeddings block. The two embeddings are jointly learned during training and summed vectors are fed to the next Transformer layers. When an email contains a HTML body without a plain text body, we extract the plain text using an HTML parser and a simple regular expression. The HTML parsing also prevents the risks of HTML obfuscation attacks which insert random HTML tags between words. 

\begin{figure}
  \centering
  \includegraphics[scale=0.6]{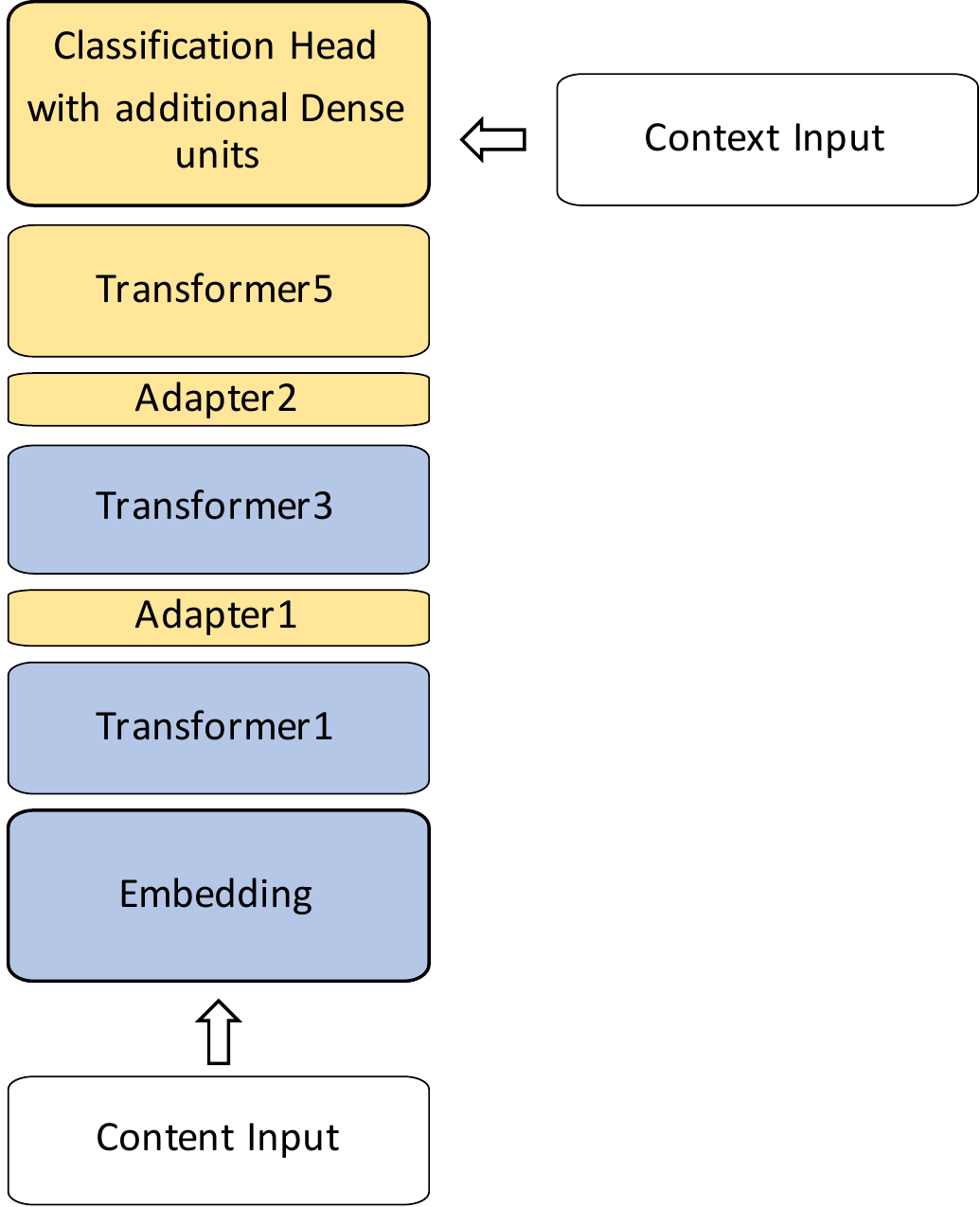}
  \caption{The architecture of context-aware BERT accepts content and context features. Our partial fine-tuning method only updates upper yellow blocks with frozen lower blue blocks.}
  \label{fig:content_context_inputs}
\end{figure}

\subsection{Context features}
Standard BERT models only accept text as input. However, an email consists of text content and header fields which provide valuable context information. Our context-aware approach combines the content from text and the context from header fields as input data. Text data is a concatenated sentence from the email’s subject and body text. We extract the following four context features from header fields. While the contextual features are quite simple and generic, they capture essential context information about the communication between senders and recipients. 

\begin{itemize}
\item Internal communication: if the sender and receiver of the communication belong to the same domain, it is internal communication.   
\item External communication: if the domain of the sender and receiver are different, it is external communication. 
\item Number of recipients: indicates the size of recipients. 
\item Number of CC (Carbon Copy): indicates the size of CC. 
\end{itemize}

To accommodate the context and content features, our model receives two inputs. As shown in figure \ref{fig:content_context_inputs}, content features are fed into the embedding layer and context features are fed into the classification layer. There are additional dense units after the last Transformer layer and the context features are combined in the dense layer with Transformer’s rich text representation, which leads to significant performance improvement. The classification head of CatBERT returns sigmoid outputs which indicate the maliciousness of input emails. 

\section{Experiments}
\label{sec:experiments}

We collected a dataset of about five million emails and associated metadata from a threat intelligence feed and internal systems at Sophos. The dataset consists of 407,161 malicious emails and 3,842,772 benign emails which were randomly sampled from a large email collection. We split the samples into training, validation and test datasets based on the first seen time to prevent data leakage issues. 70\%, 15\% and 15\% of samples were used for training, validation and testing respectively. The training dataset has 285,021 malicious and 2,697,499 benign samples, and the test dataset has 122,140 benign and 1,145,273 benign samples. For all experiments, we extracted word tokens with a max token length of 128 for email text data. 

We trained neural network models using the Pytorch framework using the Adam optimizer and 128 sized mini batches. Our implementation of DistilBERT is based on Huggingface \cite{Wolf2019HuggingFace}. We also trained a non-neural network model, Logistic regression using the scikit-learn \cite{Pedregosa2011scikit} framework. Each neural network model uses the same Multilingual BERT tokenizer for content features and uses the same size embedding layer, which is designed for each model to have similar learning capability to handle token inputs. The models are trained for 5 epochs, which was enough for the results to converge.

\subsection{Classification performance}
We first conducted experiments to compare our proposed model with two non-BERT models, Long Short-Term Memory (LSTM) \cite{hochreiter1997lstm} and Logistic Regression (LR) \cite{peng2002an}. All BERT-based models were trained with Adam optimizer and a balanced batch size of 128. Because the number of malicious samples is smaller than benign ones, we use balanced batch sampling for the imbalanced dataset. The balanced batch allocates the first 64 samples from legitimate samples and the remaining 64 samples from malicious ones.  

CatBERT, our proposed model, has three Transformer blocks with content and context features. DistilBERT contains six Transformer blocks which were used in DistilBERT paper. The LSTM model accepts word sequences as input and contains a single LSTM layer with an embedding layer which has the same 768 dimensions as BERT’s embedding layer. The neural network models were trained with Adam optimizer and a balanced batch size of 128. The LR model uses TF-IDF features from uni/bi-gram words.  

\begin{figure}
     \centering
     \begin{subfigure}[b]{0.45\textwidth}
         \centering
         \includegraphics[width=\textwidth]{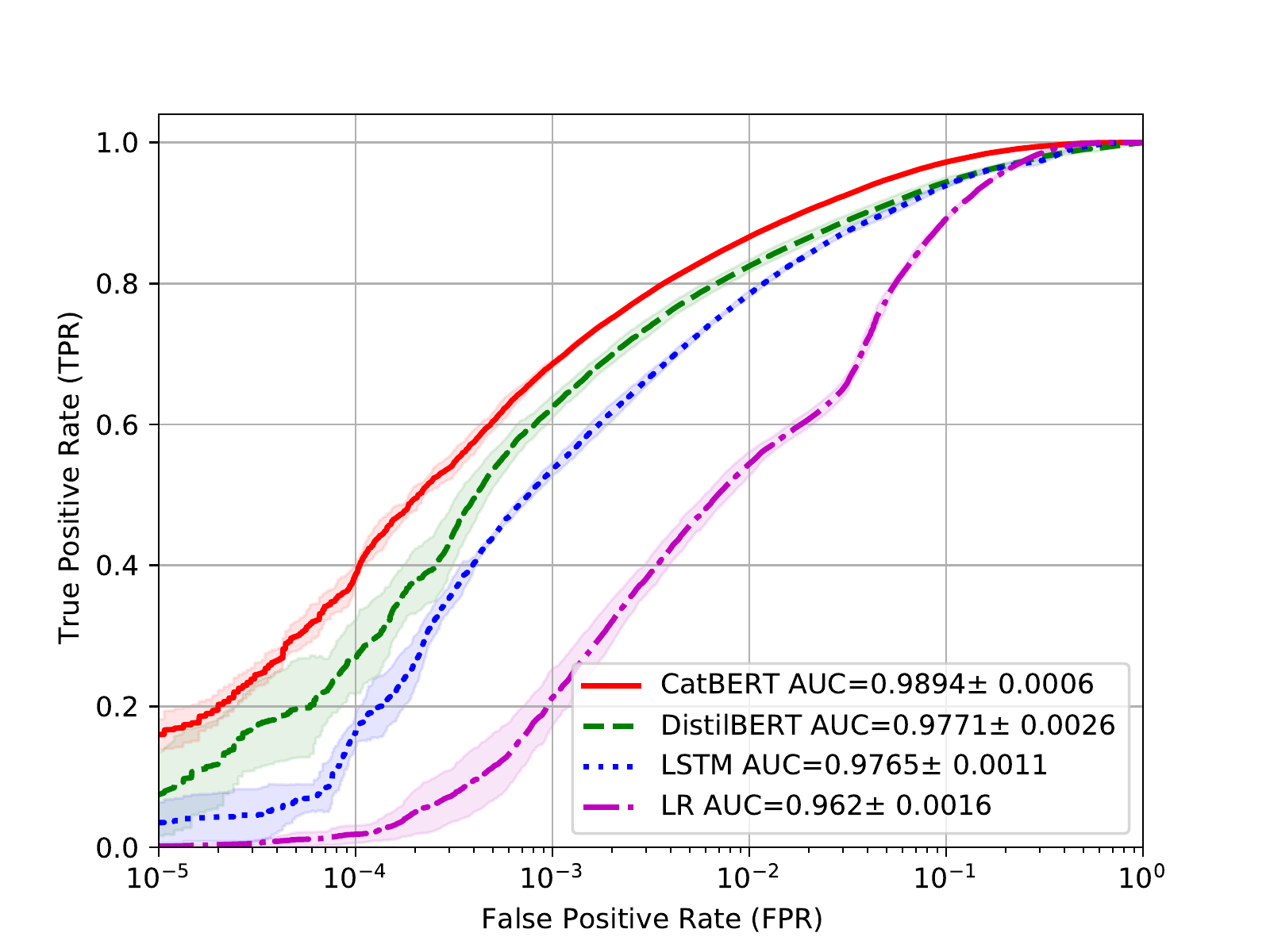}
         \caption{All samples}
     \end{subfigure}
     \begin{subfigure}[b]{0.45\textwidth}
         \centering
         \includegraphics[width=\textwidth]{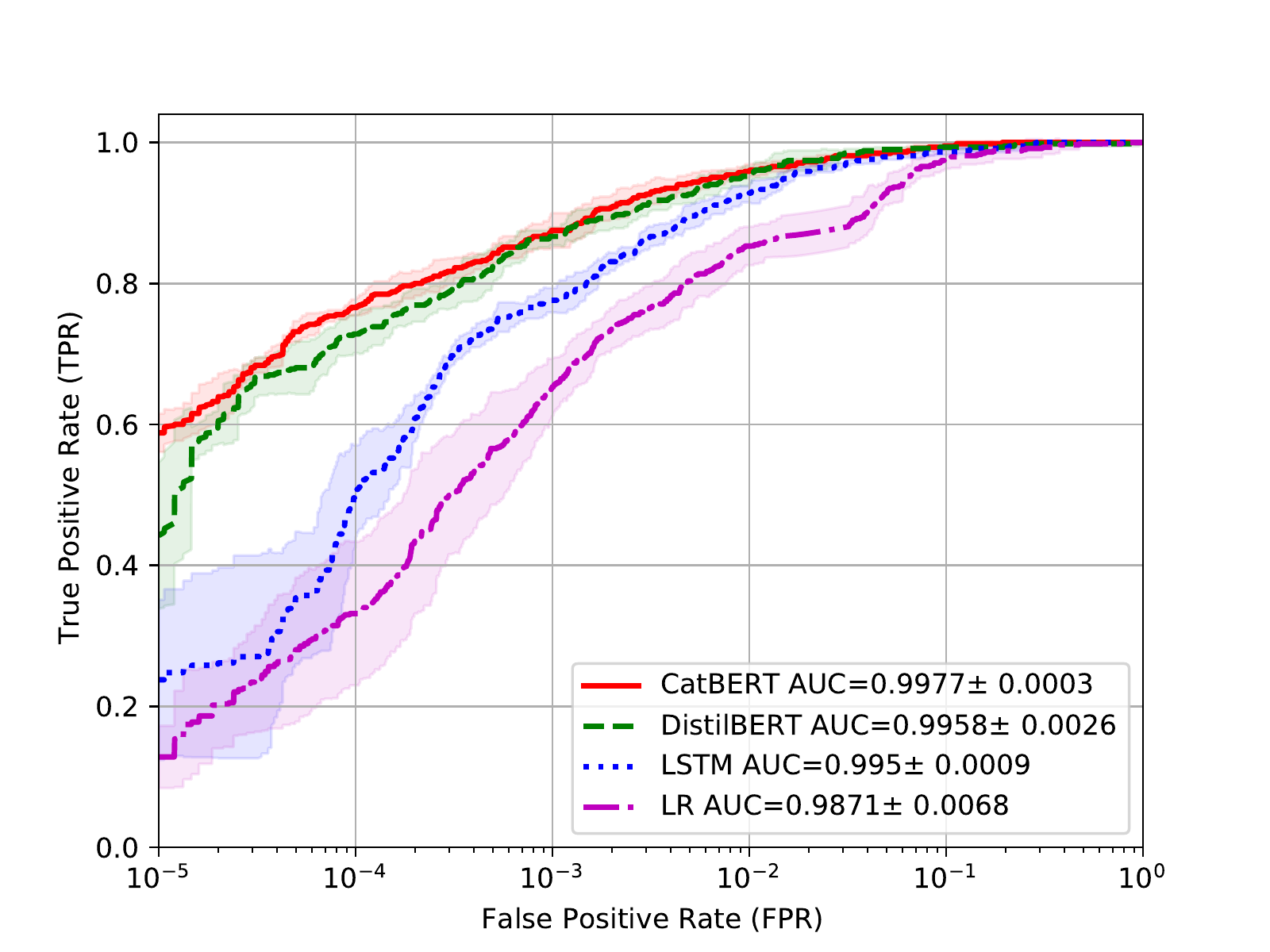}
         \caption{BEC samples}
     \end{subfigure}
     \begin{subfigure}[b]{0.45\textwidth}
         \centering
         \includegraphics[width=\textwidth]{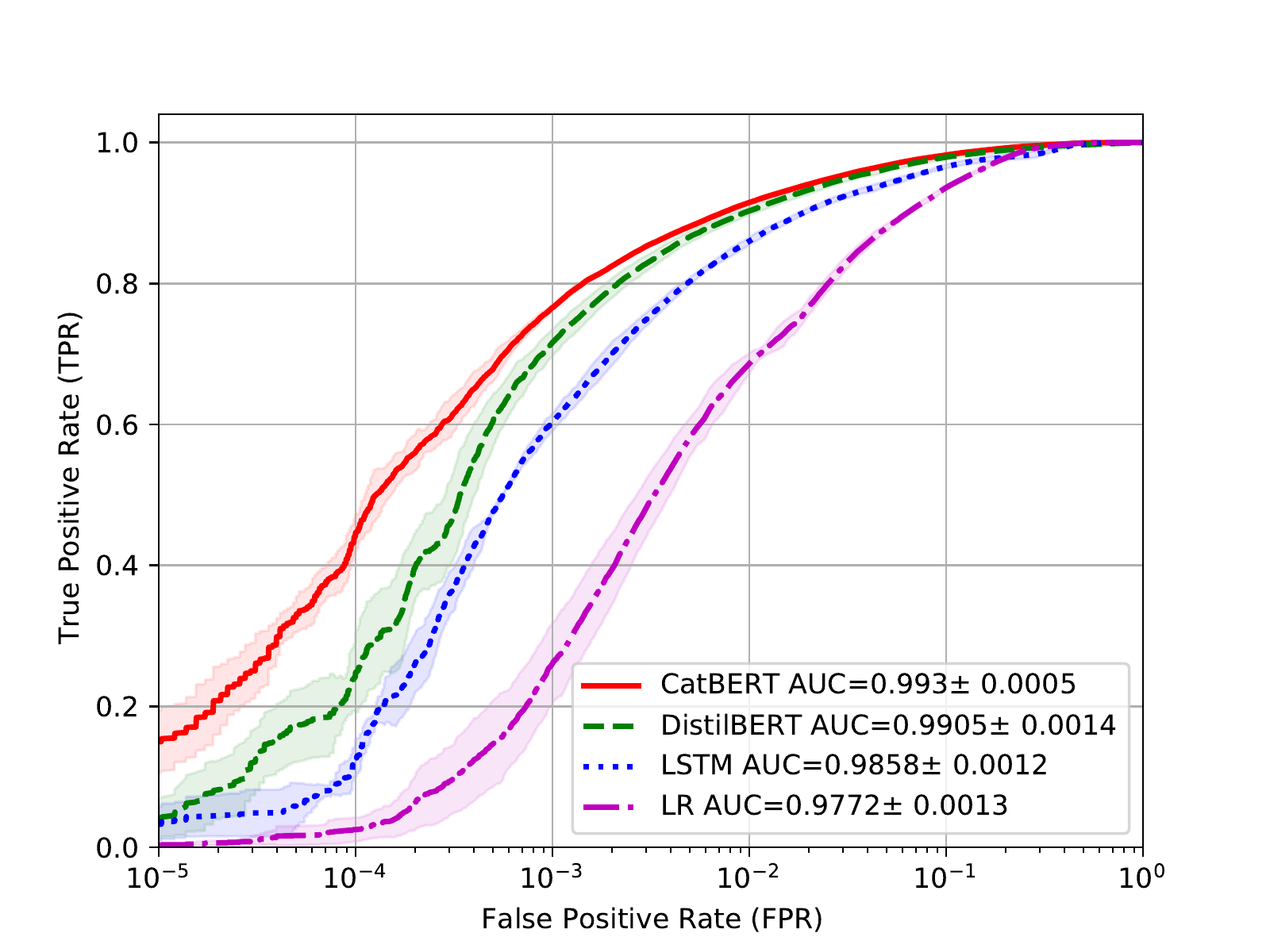}
         \caption{English samples}
     \end{subfigure}
     \begin{subfigure}[b]{0.45\textwidth}
         \centering
         \includegraphics[width=\textwidth]{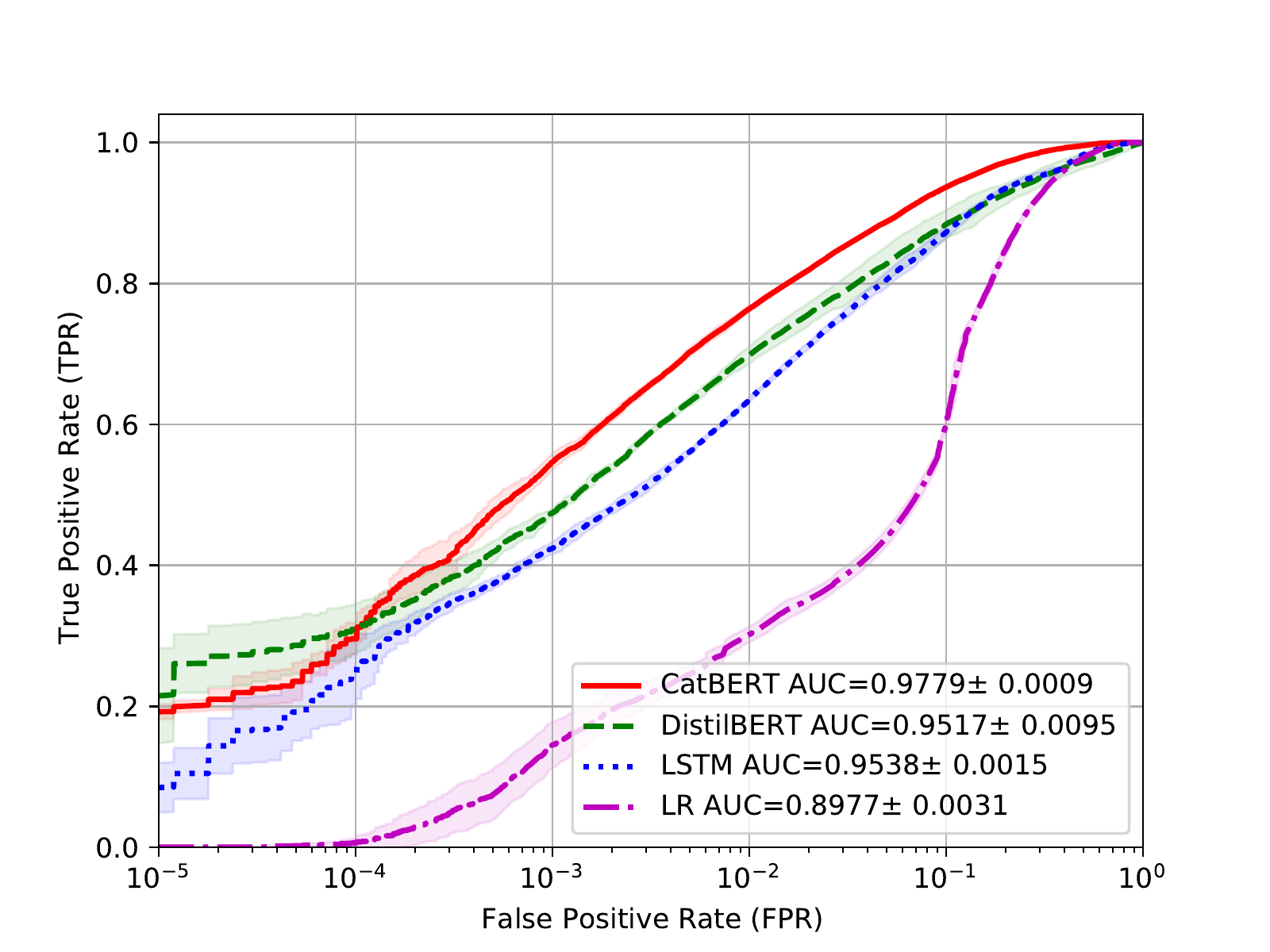}
         \caption{Non-English samples}
     \end{subfigure}
\caption{Mean ROC curves and standard deviation for CatBERT and baseline models. Top left compares four models with all test samples. Top right, bottom left, and bottom right compares only with test BEC, English and Non-English samples respectively. Mean and standard deviation are computed over five runs.}
\label{fig:fig_4modes_roc_curves}
\end{figure}

Figure \ref{fig:fig_4modes_roc_curves} compares the ROC (Receiver Operating Characteristic) curves which allow us to evaluate how the four models performs at different false positive rates. The results demonstrate that BERT-based models outperform non-BERT models by a large margin and CatBERT achieves the best performance. We divide emails samples into three groups, BEC (Business Email Compromise) samples, English samples and non-English samples. The top right shows the ROC curves for all test samples and top left is for BEC, bottom left is for English samples and bottom right is for non-English samples. Our proposed model outperformed baseline models for all four cases. Our phishing samples included BEC samples which are hand-crafted social engineering emails. To improve the detection rate for the BEC samples, we assigned a sample weight of 100 for the target samples. The method allows our model to focus on those samples, as high errors will be given when those samples are miss-classified during training.  

We compared the TPR (True Positive Rate) at different FPRs (False Positive Rates) and AUC (Area Under the Curve) for four models. Table \ref{tab:table_tpr_all_samples} and Table \ref{tab:table_tpr_bec_samples} shows the results for all test samples and test BEC samples respectively. CatBERT yielded the best performance over three baseline models.

\begin{table}
 \caption{AUC and TPRs on the all test samples for four FPRs. Mean and standard deviation results are aggregated over five runs and best ones are shown in bold.}
  \centering
  \begin{tabular}{cccccc}
    \toprule
    & \multirow{2}{*}{AUC} & \multicolumn{4}{c}{FPR} \\
    \cmidrule(r){3-6}
     &    & 0.0001 & 0.001 & 0.01 & 0.1 \\
    \midrule
    CATBERT & \textbf{0.9894 $\pm$ 0.0006} & \textbf{0.3884 $\pm$ 0.0129} & \textbf{0.6859 $\pm$ 0.0054} & \textbf{0.8663 $\pm$ 0.0017} & \textbf{0.9721 $\pm$ 0.0021} \\
    DistilBERT & 0.9771 $\pm$ 0.0026 & 0.2696 $\pm$ 0.0524 & 0.6245 $\pm$ 0.0156 & 0.8251 $\pm$ 0.0086 & 0.9436 $\pm$ 0.0066 \\
    LSTM & 0.9765 $\pm$ 0.0011 & 0.1674 $\pm$ 0.0239 & 0.5360 $\pm$ 0.0096 & 0.7852 $\pm$ 0.0048 & 0.9390 $\pm$ 0.0033 \\
    LR & 0.9620 $\pm$ 0.0016 & 0.0186 $\pm$ 0.0125 & 0.2127 $\pm$ 0.0406 & 0.5434 $\pm$ 0.0165 & 0.8918 $\pm$ 0.0045 \\
    \bottomrule
  \end{tabular}
  \label{tab:table_tpr_all_samples}
\end{table}

\begin{table}
 \caption{AUC and TPRs on the BEC samples for four FPRs. Mean and standard deviation results are aggregated over five runs and best ones are shown in bold.}
  \centering
  \begin{tabular}{cccccc}
    \toprule
    & \multirow{2}{*}{AUC} & \multicolumn{4}{c}{FPR} \\
    \cmidrule(r){3-6}
     &    & 0.0001 & 0.001 & 0.01 & 0.1 \\
    \midrule
    CATBERT & \textbf{0.9977 $\pm$ 0.0003} & \textbf{0.7658 $\pm$ 0.0116} & \textbf{0.8752 $\pm$ 0.0245} & \textbf{0.9590 $\pm$ 0.0100} & \textbf{0.9949 $\pm$ 0.0042} \\
    DistilBERT & 0.9958 $\pm$ 0.0026 & 0.7282 $\pm$ 0.0272 & 0.8667 $\pm$ 0.0128 & 0.9521 $\pm$ 0.0139 & 0.9932 $\pm$ 0.0034 \\
    LSTM & 0.9950 $\pm$ 0.0009 & 0.5060 $\pm$ 0.0652 & 0.7761 $\pm$ 0.0174 & 0.9282 $\pm$ 0.0139 & 0.9863 $\pm$ 0.0042 \\
    LR & 0.9848 $\pm$ 0.0096 & 0.2821 $\pm$ 0.1590 & 0.7077 $\pm$ 0.0598 & 0.8342 $\pm$ 0.0207 & 0.9863 $\pm$ 0.0116 \\
    \bottomrule
  \end{tabular}
  \label{tab:table_tpr_bec_samples}
\end{table}

\begin{figure}
  \centering
  \includegraphics[scale=0.6]{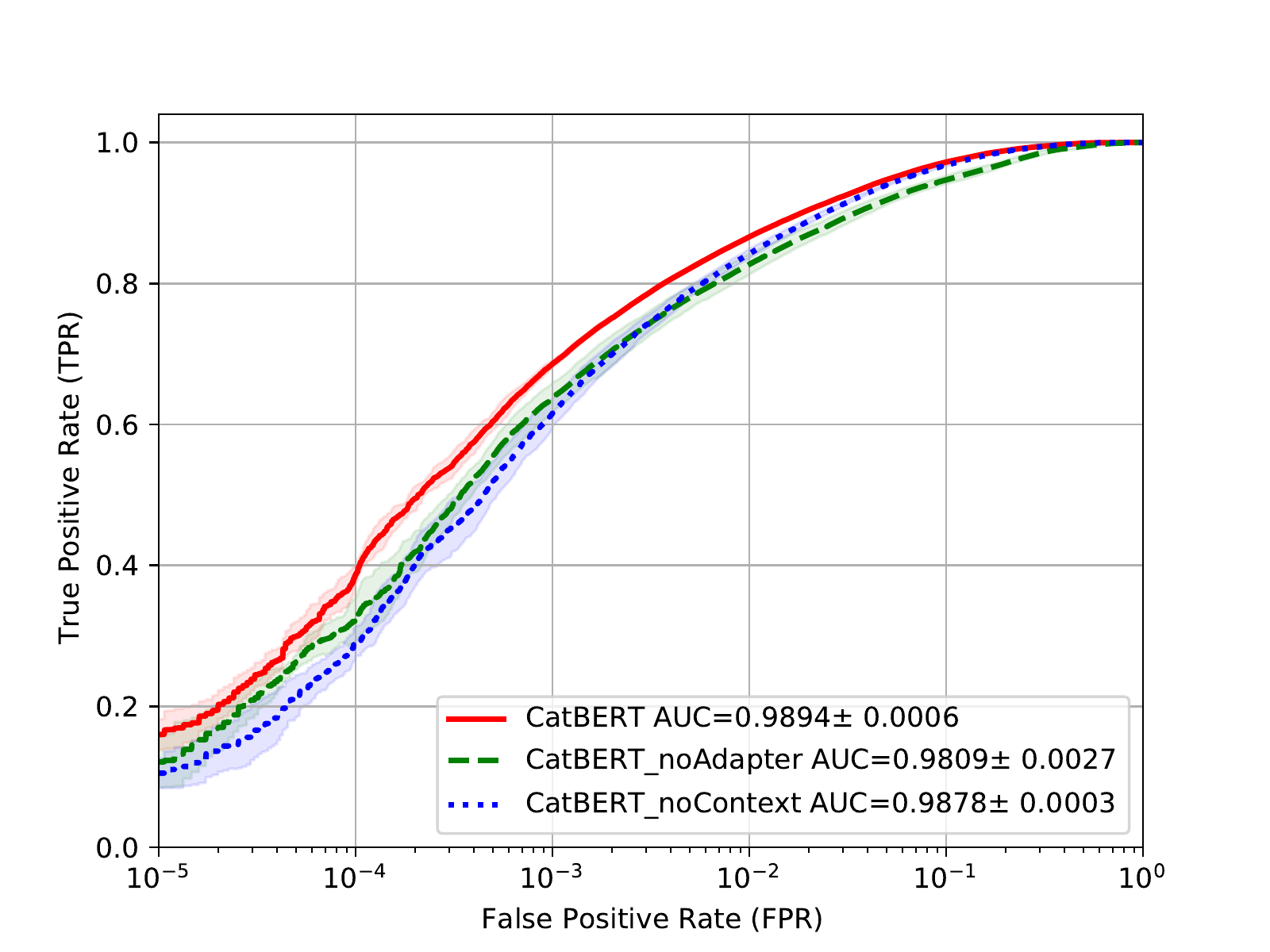}
  \caption{The ROC curves for ablation study. Red, green and blue line shows CatBERT, CatBERT without adapters and CatBERT without context input respectively. Mean and standard deviation are computed over five runs.}
  \label{fig:email_catbert_ablation_rocs}
\end{figure}

We also conducted an ablation study to investigate the influence of additional components in CatBERT. Figure \ref{fig:email_catbert_ablation_rocs} compares the CatBERT model with two modified models. The CatBERT\_noAdapter in green colour has no adapter layers and the CatBERT\_noContext in blue colour has no context related layers. When one of the adapter or context layers was removed from CatBERT, there was a measurable performance drop.

\subsection{Runtime performance}
One of the main obstacles in applying full-sized Transformer-based models to real-time detection systems is the runtime performance of the full models. As millions of emails need to be processed daily in a production environment, the inference speed of any machine learning model that will be analyzing emails is a critical performance metric for model deployment. 

Table \ref{tab:table_model_size} compares the model size and inference speed of neural network models. DistilBERT has 135 million parameters, which is the largest model with 6 Transformer blocks, and a correspondingly long inference time. CatBERT with 3 Transformer blocks has 117 million parameters, is about 15\% smaller than DistilBERT, and obtains 1.6x speed up in CPU inference time (using an AWS m5.large instance type) and 1.3x speed up on GPU (AWS g3s.xlarge instance type). While the LSTM model obtains the fastest inference speed in CPU inference, having only 96 million parameters, it performs the worst with respect to AUC. The LSTM model runs slower than CatBERT on GPU as recurrent neural networks do not run in parallel. The relatively modest reduction in parameter size for CatBERT is due to the fact that all three models use a large embedding layer with 92 million parameters, which accounts for about 70\% of all parameters. When we consider the results of detection performance and inference speed, CatBERT achieved the best performance. 

\newcolumntype{L}{>{\centering\arraybackslash}m{1.5cm}}

\begin{table}
 \caption{Comparison of model size and inference speed, CatBERT achieved the best AUC and inference speed.}
  \centering
  \begin{tabular}{LLLLLLLL}
    \toprule
     & Number of Transformers & Number of total parameters (millions) & Number of embedding parameters \newline (millions) & Number of non-embedding parameters (millions) & Inference time on CPU (milliseconds) & Inference time on GPU & AUC  \\
    \midrule
    DistilBERT & 6 & 135 (1.2x) & 92 & 43 (1.7x) & 130 (1.6x) & 9.4 (1.3x) & 0.9771 \\
    CatBERT & 3 & 117 (1x) & 92 & 25 (1x) & 79 (1x) & \textbf{7.0} (1x) & \textbf{0.9894} \\
    LSTM & N/A & \textbf{96} (0.8x) & 92 & 4 (0.2x) & \textbf{69} (0.9x) & 14.7 (2x) & 0.9765 \\
    \bottomrule
  \end{tabular}
  \label{tab:table_model_size}
\end{table}

\subsection{Robustness to adversarial attacks}
Adversaries often deform malicious emails using deliberate typos and synonyms to circumvent machine learning models, as machine learning models are known to be vulnerable to adversarial actions \cite{bhowmick2016machine}. Our approach is less prone to the adversarial attacks as our model has been pre-trained with a large text dataset, and also understands the main intent of emails even if some of words are replaced with typographic errors or synonyms. In addition, our context features from email header fields contribute to CatBERT’s resilience to those attacks. 

\begin{figure}
     \centering
     \begin{subfigure}[b]{0.8\textwidth}
         \centering
         \includegraphics[width=\textwidth]{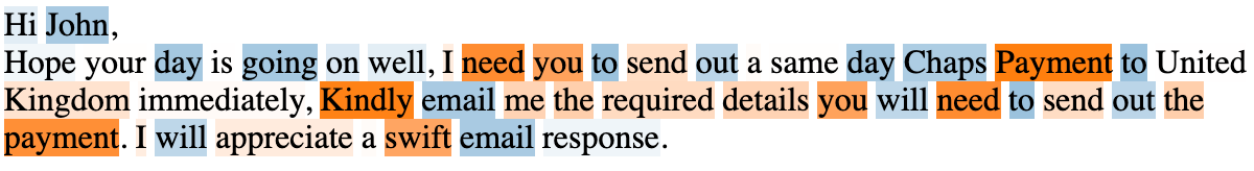}
         \caption{Original sample }
     \end{subfigure}
     \begin{subfigure}[b]{0.8\textwidth}
         \centering
         \includegraphics[width=\textwidth]{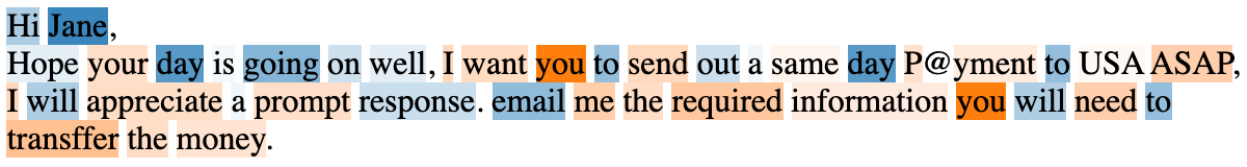}
         \caption{Modified sample}
     \end{subfigure}
\caption{Model’s predictions can be examined by highlighted words using LIME. Top: original sample, bottom: modified sample by adding typos, replacing words with synonyms and removing some words. Our model detects the modified sample by identifying relevant tokens.}
\label{fig:fig_lime_analysis}
\end{figure}

To inspect our model’s robustness, we examine CatBERT with LIME \cite{ribeiro2016why} which explains models by learning an interpretable model locally around the prediction. BERT models are complex and hard to interpret, whereas LIME learns a linear regression model with neighborhood samples and provides model’s interpretability with learned weights, albeit only for a single specific sample. The individual tokens may then be interpreted by examining their impact on the prediction probabilities for the local linear model. For example, in figure \ref{fig:fig_lime_analysis}, the orange color indicates suspicious words with high positive weights, while the blue color indicates benign ones. The payment word has the highest weight and is highlighted in the email text. 
The analysis with LIME demonstrates that our model can identify suspicious signals in the email content. The top example shows the LIME’s weights for an original sample, and the bottom one shows the weights for a modified sample. The modified sample was generated by adding typos, replacing words with synonyms, removing some words or changing word sequence. As shown in table \ref{tab:suspicious_tokens} which lists high-weighted suspicious tokens from the original and modified sample, our model was able to detect the modified sample by identifying relevant tokens. Unicode attack is often employed by attackers to obfuscate keywords with Unicode characters \cite{fu2006the}. For example, the “Payment” word in the original sample was replaced by “P@yment”. However, our model was able to identify the word as “P” and “yment” as suspicious ones. BERT tokenizer divides complex words into sub words and the two tokens are listed in the high-weighted tokens.

\begin{table}
 \caption{Identified suspicious tokens from the original and modified sample.}
  \centering
  \begin{tabular}{cc}
    \toprule
    & High-weighted tokens  \\
    \midrule
    Original sample & Payment, payment, Kindly, need, you, swift, details, required, send, appreciate \\
    Modified sample & you, transffer, required, me, need, ASAP, send, want, prompt, P, yment, money \\
    \bottomrule
  \end{tabular}
  \label{tab:suspicious_tokens}
\end{table}

\begin{figure}
     \centering
     \begin{subfigure}[b]{0.45\textwidth}
         \centering
         \includegraphics[width=\textwidth]{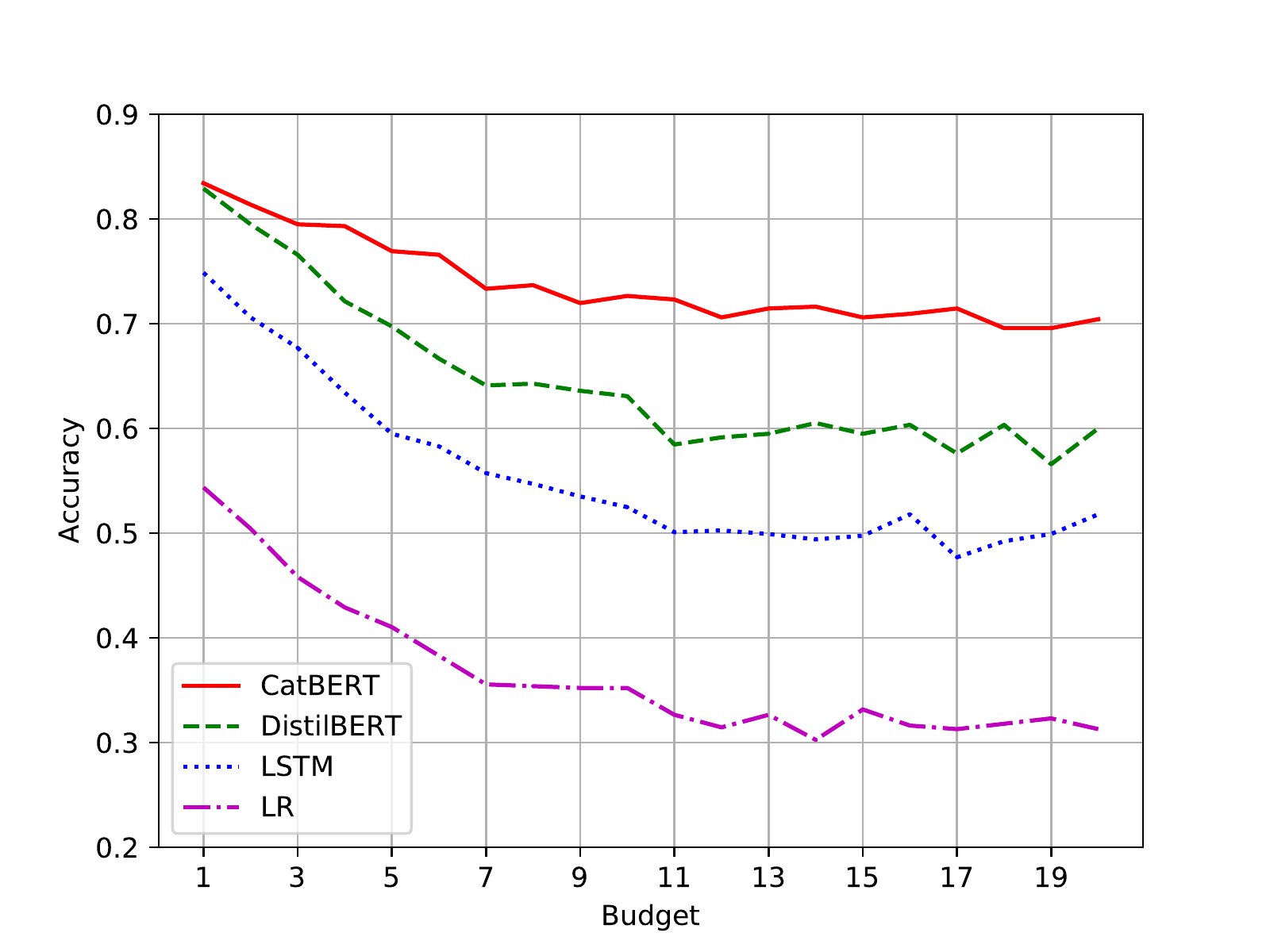}
         \caption{Synonyms attacks}
     \end{subfigure}
     \begin{subfigure}[b]{0.45\textwidth}
         \centering
         \includegraphics[width=\textwidth]{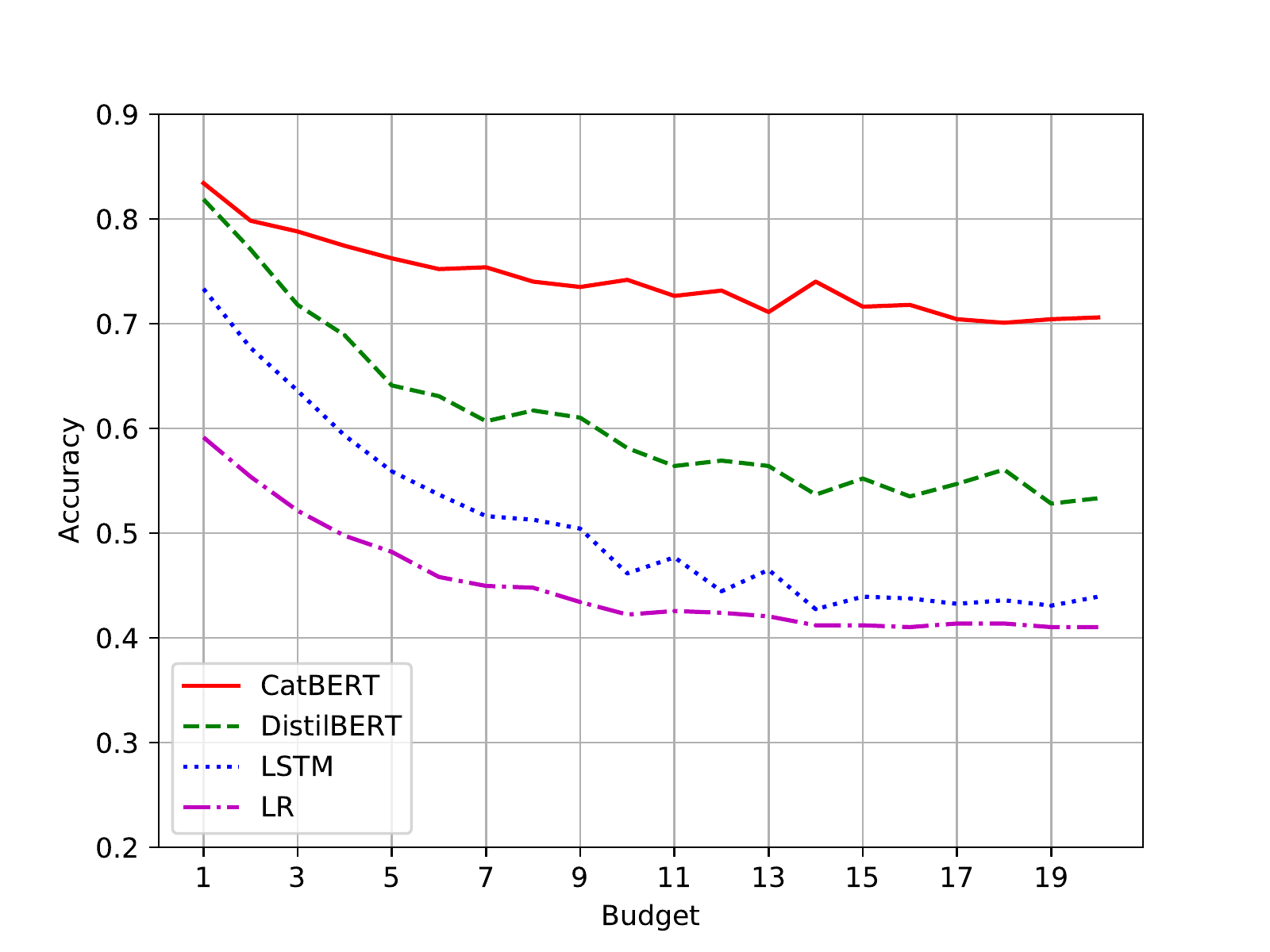}
         \caption{Typo attacks}
     \end{subfigure}
\caption{The mean detection accuracy against adversarial attacks over five runs. Left shows the accuracy against synonyms attacks and right shows the accuracy against typo attacks. CatBERT is more robust than baseline models.}
\label{fig:fig_adversarial_attack}
\end{figure}

To further analyze the model’s robustness, we created adversarial samples from the BEC samples in our test dataset using simple synonyms and typos. The modified samples were generated by replacing a word with its synonym or typo. Figure \ref{fig:fig_adversarial_attack} compares the accuracy for CatBERT and baseline models with the adversarial samples. The left plot shows that results for synonym attack which replaces words with synonyms. The right plot shows the results for typo attack which modifies words with typos. While CatBERT obtained the best accuracies for both attacks, baseline models produced accuracies which were significantly degraded by the attacks, which suggests that CatBERT is more robust to such adversarial attacks.

\section{Conclusion}
\label{sec:conclusion}
Hand-crafted social engineering emails pose a significant challenge for traditional signature and ML detection technologies, as an individually targeted email may not share word sequences or word choices with previously seen attacks. We introduce an efficiently downsized BERT model by fine-tuning a pre-trained, highly pruned BERT model with additional context features from email headers that can detect targeted phishing emails, even in the presence of deliberate misspellings and attempted evasions. Our approach outperforms strong baseline models with adapter and context layers. CatBERT is 15\% smaller and 160\% faster than DistilBERT which is already 40\% smaller than vanilla BERT. Our model is also resilient to randomly introduced changes in word choice, word ordering and other permutations.

\bibliographystyle{unsrt}  


\begin{thebibliography}{1}

\bibitem{devlin2018bert}
Jacob Devlin, Ming-Wei Chang, Kenton Lee, and Kristina Toutanova.
\newblock Bert: Pre-training of deep bidirectional transformers for language understanding.
\newblock In {\em arXiv preprint arXiv:1810.04805}, 2018.

\bibitem{sanh2019distilbert}
Victor Sanh, Lysandre Debut, Julien Chaumond, Thomas Wolf.
\newblock DistilBERT, a distilled version of BERT: smaller, faster, cheaper and lighter.
\newblock In {\em arXiv preprint arXiv:1910.01108}, 2019.

\bibitem{ribeiro2016why}
Marco Tulio Ribeiro, Sameer Singh, Carlos Guestrin.
\newblock "Why Should I Trust You?": Explaining the Predictions of Any Classifier.
\newblock In {\em 97-101. 10.18653/v1/N16-3020}, 2016.

\bibitem{lan2019albert}
Zhenzhong Lan, Mingda Chen, Sebastian Goodman, Kevin Gimpel, Piyush Sharma, Radu Soricut.
\newblock ALBERT: A Lite BERT for Self-supervised Learning of Language Representations.
\newblock In {\em arXiv preprint arXiv:1909.11942}, 2019.

\bibitem{jiao2019tinybert}
Xiaoqi Jiao, Yichun Yin, Lifeng Shang, Xin Jiang, Xiao Chen, Linlin Li, Fang Wang, Qun Liu.
\newblock TinyBERT: Distilling BERT for Natural Language Understanding.
\newblock In {\em arXiv preprint arXiv:1909.10351}, 2019.

\bibitem{lukas2020catching}
Halgaš, Lukáš \& Agrafiotis, Ioannis \& Nurse, Jason.
\newblock Catching the Phish: Detecting Phishing Attacks Using Recurrent Neural Networks (RNNs).
\newblock In {\em Information Security Applications, pages 219-233}, 2020.

\bibitem{hochreiter1997lstm}
Hochreiter, S., Schmidhuber, J.
\newblock Long short-term memory.
\newblock In {\em Neural Computation 9(8), 1735–1780}, 1997.

\bibitem{verma2012detecting}
Verma, R., Shashidhar, N., Hossain, N.
\newblock Detecting phishing emails the natural language way.
\newblock In {\em 17th European Symposium on Research in Computer Security. pp. 824–841}, 2012.

\bibitem{nikita2020deepquarantine}
Benkovich, Nikita \& Dedenok, Roman \& Golubev, Dmitry.
\newblock DeepQuarantine for Suspicious Mail.
\newblock In {\em arXiv preprint arXiv:2001.04168}, 2020.

\bibitem{ho2019detecting}
Ho, Grant \& Cidon, Asaf \& Gavish, Lior \& Schweighauser, Marco \& Paxson, Vern \& Savage, Stefan \& Voelker, Geoffrey \& Wagner, David.
\newblock Detecting and Characterizing Lateral Phishing at Scale.
\newblock In {\em Proceedings of the 28th USENIX Conference on Security Symposium}, 2019.

\bibitem{ho2019improved}
Andre Bergholz, Jeong Ho Chang, Gerhard Paaß, Frank Reichartz, and Siehyun Strobel.
\newblock Improved Phishing Detection using Model-Based Features.
\newblock In {\em Proc. of 5th CEAS}, 2008.

\bibitem{radford2019language}
Alec Radford, Jeff Wu, Rewon Child, David Luan, Dario Amodei, and Ilya Sutskever.
\newblock Language models are unsupervised multitask learners.
\newblock 2019.

\bibitem{hinton2015distilling}
Geoffrey Hinton, Oriol Vinyals, and Jeff Dean.
\newblock Distilling the knowledge in a neural network.
\newblock In {\em arXiv preprint arXiv:1503.02531}, 2015.

\bibitem{han2015deep}
Song Han, Huizi Mao, and William J Dally.
\newblock Deep compression: Compressing deep neural networks with pruning, trained quantization and huffman coding.
\newblock In {\em arXiv preprint arXiv:1510.00149}, 2015.

\bibitem{fu2006the}
Fu, Anthony \& Deng, Xiaotie \& Wenyin, Liu \& Little, Greg.
\newblock The methodology and an application to fight against Unicode attacks.
\newblock In {\em SOUPS}, 2006.

\bibitem{bhowmick2016machine}
Bhowmick, Alexy \& Hazarika, Shyamanta.
\newblock Machine Learning for E-mail Spam Filtering: Review,Techniques and Trends.
\newblock In {\em arXiv preprint arXiv:1606.01042}, 2016.

\bibitem{ho2017detecting}
Grant Ho and Aashish Sharma and Mobin Javed and Vern Paxson and David Wagner.
\newblock Detecting Credential Spearphishing in Enterprise Settings.
\newblock In {\em USENIX Security Symposium}, 2017.

\bibitem{cidon2019high}
Asaf Cidon and Lior Gavish, Itay Bleier, Nadia Korshun, Marco Schweighauser and Alexey Tsitkin.
\newblock High Precision Detection of Business Email Compromise.
\newblock In {\em USENIX Security Symposium}, 2019.

\bibitem{houlsby2019parameter}
Neil Houlsby, Andrei Giurgiu, Stanislaw Jastrzebski, Bruna Morrone, Quentin de Laroussilhe, Andrea Gesmundo, Mona Attariyan, Sylvain Gelly.
\newblock Parameter-Efficient Transfer Learning for NLP.
\newblock In {\em ICML}, 2019.

\bibitem{Pedregosa2011scikit}
Fabian Pedregosa and Ga{{\"e}}l Varoquaux and Alexandre Gramfort and Vincent Michel and Bertrand Thirion and Olivier Grisel and Mathieu Blondel and Peter Prettenhofer and Ron Weiss and Vincent Dubourg and Jake Vanderplas and Alexandre Passos and David Cournapeau and Matthieu Brucher and Matthieu Perrot and {{\'E}}douard Duchesnay
\newblock Scikit-learn: Machine Learning in Python.
\newblock In {\em JMLR 12, pp. 2825-2830}, 2011.

\bibitem{Wolf2019HuggingFace}
Thomas Wolf, Lysandre Debut, Victor Sanh, Julien Chaumond, Clement Delangue, Anthony Moi, Pierric Cistac, Tim Rault, Rémi Louf, Morgan Funtowicz, Joe Davison, Sam Shleifer, Patrick von Platen, Clara Ma, Yacine Jernite, Julien Plu, Canwen Xu, Teven Le Scao, Sylvain Gugger, Mariama Drame, Quentin Lhoest, Alexander M. Rush
\newblock HuggingFace's Transformers: State-of-the-art Natural Language Processing.
\newblock In {\em arXiv preprint arXiv:1910.03771}, 2019.

\bibitem{peng2002an}
Peng, Joanne \& Lee, Kuk \& Ingersoll, Gary.
\newblock An Introduction to Logistic Regression Analysis and Reporting.
\newblock In {\em Journal of Educational Research}, 2002.

\end{thebibliography}

\end{document}